\documentclass[10pt,twocolumn,letterpaper]{article}

\usepackage{cvpr}
\usepackage{times}
\usepackage{epsfig}
\usepackage{graphicx}
\usepackage{amsmath}
\usepackage{amssymb}
\usepackage{microtype}
\usepackage{enumitem}
\usepackage{subcaption}
\usepackage{algorithm}
\usepackage{algpseudocode}
\usepackage{makecell}
\usepackage[dvipsnames]{xcolor}
\usepackage{rotating}

\usepackage[pagebackref=true,breaklinks=true,letterpaper=true,colorlinks,bookmarks=false]{hyperref}

\cvprfinalcopy 

\def\cvprPaperID{6431} 

\ifcvprfinal\fi
\begin{document}

\title{Constrained Generative Adversarial Networks for Interactive Image Generation}

\author{Eric Heim\\
  Air Force Research Laboratory\\
  Information Directorate \\
Rome, NY USA\\
{\tt\small eheim602@gmail.com}}

\renewcommand{\algorithmicrequire}{\textbf{Input:}}
\renewcommand{\algorithmicensure}{\textbf{Output:}}
\algdef{S}{Hyperparameters}[1]{\textbf{Hyperparameters: } \mbox{#1}}%
\algrenewcommand\algorithmicindent{0.5em}

\maketitle

\begin{abstract}
  Generative Adversarial Networks (GANs) have received a great deal of attention due in part to recent success in generating original, high-quality samples from visual domains.
  However, most current methods only allow for users to guide this image generation process through limited interactions.
  In this work we develop a novel GAN framework that allows humans to be ``in-the-loop'' of the image generation process.
  Our technique iteratively accepts relative constraints of the form ``Generate an image more like image $A$ than image $B$''.
  After each constraint is given, the user is presented with new outputs from the GAN, informing the next round of feedback.
  This feedback is used to constrain the output of the GAN with respect to an underlying semantic space that can be designed to model a variety of different notions of similarity (e.g. classes, attributes, object relationships, color, etc.).
  In our experiments, we show that our GAN framework is able to generate images that are of comparable quality to equivalent unsupervised GANs while satisfying a large number of the constraints provided by users, effectively changing a GAN into one that allows users interactive control over image generation without sacrificing image quality.
\end{abstract}

\section{Introduction}
Learning a generative model from data is a task that has gotten recent attention due to a number of breakthroughs in complex data domains~\cite{kingma2014auto,van2016pixel,karras2018progressive}.
Some of the most striking successes have been in creating novel imagery using Generative Adversarial Networks (GANs)~\cite{goodfellow2014generative}.
While GANs show promise in having machines effectively ``draw'' realistic pictures, the mechanisms for allowing humans to guide the image generation process have been largely limited to conditioning on class labels~\cite{mirza2014conditional} (e.g. ``Draw a zero.'') or domain-specific attributes \cite{yan2016attribute2image} (e.g. ``Draw a coat with stripes.''). 
Such feedback, though powerful, limits the user to expressing feedback through a pre-defined set of labels.
If the user is unable to accurately express the characteristics that they desire using this label set, then they cannot guide the model to produce acceptable images.

%
\begin{figure}[t]
\begin{center}
   \includegraphics[width=0.9\linewidth]{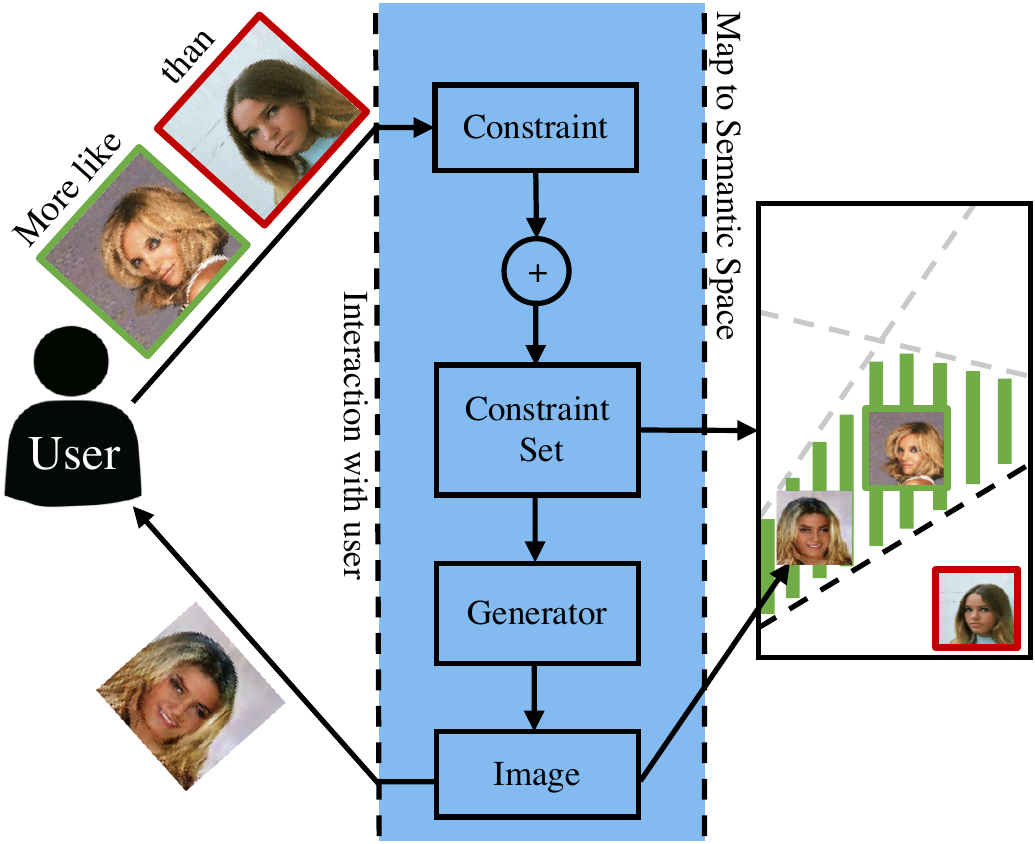}
\end{center}
\caption{Interaction with the CONGAN generator: A user provides a relative constraint in the form of two images meaning ``Generate an image more like image $A$ than image $B$.'' The constraint is combined to previously given constraints to form a set, which is input to the generator to produce an image. This image is shown to the user to drive further iterations of feedback. The goal for the generator is to ``satisfy'' the constraints with respect to a mapping to an underlying  semantic space.  The generator satisfies a constraint $\left(A,B\right)$ by producing an image that is mapped to a coordinate closer to where $A$ is mapped than to where $B$ is mapped.}
\label{fig:NotionalDiagram}
\end{figure}

In this work, we seek a more natural and powerful way for humans to interact with a generative model.
To this end, we propose a novel GAN technique we call CONstrained GAN (CONGAN).
Our model is designed to accept human feedback iteratively, effectively putting users ``in-the-loop'' of the generation process.
Figure \ref{fig:NotionalDiagram} illustrates how a user interacts with the CONGAN generator.
The generator accepts \emph{relative} constraints of the form ``More like image $A$ than image $B$.''
These constraints are used to define a feasible region within a given \emph{semantic space} that models an underlying notion of similarity between images.
The goal of the generator is to accept relative constraints as input, and output an image that is within the corresponding feasible region.

Modeling interaction in this way has two primary benefits.
First, such relative pair-wise assessments have been shown to be an easy medium for humans to articulate similarity~\cite{kendall1990rank,stewart2005absolute}.
As such, CONGAN allows users to refine its output in a natural way.
Relative constraints can also be used to allow for different interactions besides providing pair-wise comparisons.
If the output from the generator is then input as the $B$ (``less similar'') image in the next iteration, the user need only provide the $A$ (``more similar'') image.
In this way, the user can provide single examples, meaning ``More like image $A$ than what was previously generated'', to refine the output.
Second, within the CONGAN framework, the semantic space defines the characteristics that users guide the image generation process.
This allows for the option of a variety of notions of similarity, such as class and attribute information, but also more continuous or complex notions such as color spaces, or size of objects within the image.



To achieve this form of interaction, our model must have multiple interrelated components.
The generator must be able to accept a variable number of constraints as a \emph{set}, i.e. the output should be invariant to the order of the input constraints.
For this, we leverage recent work in Memory Networks~\cite{graves2014neural,weston2015memory,sukhbaatar2015end,vinyals2016order} within the CONGAN generator to learn a fixed length vector representation over the constraint set as a whole.
In addition, the generator must not only be able to generate realistic looking images, but images that are within the feasible region of a given semantic space.
During training, the CONGAN generator is trained against a \emph{constraint critic} that enforces the output to satisfy given constraints.
The result is a generator that is able to produce imagery guided by iterative relative feedback.

The remainder of the paper will proceed as follows.
First, we discuss prior related work.
Then, we describe our method beginning with a formal definition of the constrained generation problem, continuing to an outline of the CONGAN training algorithm, and ending with a description of the CONGAN generator.
Next, we perform an evaluation where we compare our method to an unsupervised GAN, showing qualitative and quantitative results.
Finally, we conclude.

\section{Related Work}
Our proposed CONGAN method follows from a long line of work in neural network image generation.
Specifically, autoencoders~\cite{kingma2014auto}, autoregressive models~\cite{van2016pixel}, and generative adversarial networks~\cite{goodfellow2014generative} (GANs) have all shown recent success.
We chose to learn a model using the GAN framework, as GANs are arguably the best performing generative models in terms of qualitative image quality.

Much of the fundamental work in GANs have focused on unsupervised learning settings~\cite{goodfellow2014generative,zhao2016energy,arjovsky2017wasserstein}.
The output of these models can be controlled by manipulating the latent space used as input~\cite{radford2015unsupervised,perarnau2016invertible}.
However, such manipulation is limited in that the latent space often has no obvious human understandable interpretation.
Thus finding ways to manipulate it requires either trial and error or interpolating between two points in the latent space.
Other works learn \emph{conditional} GAN models~\cite{mirza2014conditional}, where generation is guided by side information, such as class labels~\cite{mirza2014conditional}, visual attributes~\cite{yan2016attribute2image}, text~\cite{reed2016generative}, and images~\cite{van2016conditional}.
In this work, we aim to develop a method that allows more intuitive manipulation of a GANs output that generalizes to many different forms of similarity.

The GAN method most similar to ours is the one introduced in~\cite{zhu2016generative}.
This method first maps an image to a manifold of natural images using a GAN.
Then, they provide a series of image editing operations that users can use to move the image along that manifold.
We see our work as related but orthogonal to this work as both the means for manipulation, as well as the goals of the methods differ.

Another line of research that motivates this work is interactive learning over imagery.
Much of the work in this field has focused on classification problems ~\cite{branson2010visual,wah2011multiclass,kumar2012leafsnap,Wah_2014_CVPR}, but also others such as learning localized attributes~\cite{duan2012discovering}.
Most notably, in~\cite{kovashka2012whittlesearch} the authors propose an interactive image search method that allows users to provide iterative refinements to their query, based on visual attributes.
This is similar in principle to our method in that their method searches images through interactive comparisons to other images in the domain of interest.
However, our method does not necessarily require predefined attributes and generates novel imagery instead of retrieving relevant images from a database.

\section{A Model for Constrained Image Generation}
The goal of this work is to learn an image generation model in the form of a mapping from a set of pair-wise relative constraints to a realistic looking image.
Let $\mathcal{X}$ be a domain of images.
We wish to learn the mapping:
\begin{equation*}
g_{\Theta}:\left\{\left(\mathcal{X}{\times}\mathcal{X}\right)^i\ |\ i \geq 1\right\} \times Z \mapsto \mathcal{X}
\end{equation*}
This \emph{generator} maps a set of constraints $\mathcal{C} =\left\{C_1,C_2,...\right\}$ and a random noise vector $\mathbf{z} \in Z$ to an image, where a constraint $C=\left(\mathbf{X}_+,\mathbf{X}_-\right) \in \mathcal{X} \times \mathcal{X}$ is a pair of images meaning ``Generate an image more like $\mathbf{X}_+$ than $\mathbf{X}_-$.''
Intuitively, $\mathbf{z}$ represents the variation of imagery allowed within the constraints, and different $\mathbf{z}$ will produce different images.
Practically, $\mathbf{z}$ provides the noise component necessary for our generator to be trained within the GAN framework.

For training our generator, we require a mechanism that determines whether the output of $g_{\Theta}$ satisfies input constraints.
To this end, we assume the existence of a mapping $\phi : \mathcal{X} \mapsto \mathcal{S}$ that maps images to a \emph{semantic space}.
The only requirements are that $\phi$ be differentiable, and that there exists a distance metric $d_{\mathcal{S}}$ over elements of $\mathcal{S}$.
For instance, if one wanted to have users manipulate generated images by their attributes (i.e. the dimensions of $\mathcal{S}$ correspond to attributes), $\phi$ could be a learned attribute classifier (for binary attributes) or regressor (for continuous attributes).
We say a generated image $\hat{\mathbf{X}}$ satisfies a given constraint $C=\left(\mathbf{X}_+,\mathbf{X}_-\right)$ with respect to $\mathcal{S}$ if the following holds:
\begin{equation}
d_{\mathcal{S}}\left(\phi(\hat{\mathbf{X}}),\phi(\mathbf{X}_+)\right) < d_{\mathcal{S}}\left(\phi(\hat{\mathbf{X}}),\phi(\mathbf{X}_-)\right)
\label{eq:const_sat}
\end{equation}
Given a set of constraints $\mathcal{C}$, the goal of $g_{\Theta}$ is to produce an $\hat{\mathbf{X}}$ that satisfies all constraints in the set.
In doing so, the generator produces images that are closer in the semantic space to ``positive'' images $\mathbf{X}_+$ than ``negative'' images $\mathbf{X}_-$.
Put another way, $\mathcal{C}$ defines a feasible region in $\mathcal{S}$ for which $\hat{\mathbf{X}}$ must lie in.
How we use this idea of relative constraints to train $g_{\Theta}$ is discussed in the following section.

\subsection{Adversarial Training with Relative Constraints}
To train the generator $g_{\Theta}$, we utilize the GAN framework that pits a generator $g_{\Theta}$ against a discriminator $d_{W}$, where both $g$ and $d$ and neural networks parameterized by $\Theta$ and $W$, respectively.
The discriminator is trained to distinguish outputs of the generator from real image samples.
The generator is trained to produce images that the discriminator cannot differentiate from real samples.
The two are trained against one another; at convergence, the generator is often able to produce instances that are similar to real samples.

\begin{algorithm}[t]
  \caption{CONGAN Training Procedure}
  \label{alg:CONGAN_train}
\begin{algorithmic}
\Require{Gradient penalty coefficient $\lambda$, constraint penalty coefficient $\gamma$, discriminator iterations per generator iteration $n_{disc}$, batch size $m$, Adam optimizer parameters $\alpha,\beta_1,\beta_2$}
\Repeat
        \For{$t=1,...n_{disc}$}
                \For{$i=1,...,m$}
                        \State{Sample $\mathbf{X}\sim\mathbb{P}_{\mathcal{D}}, \mathcal{C}\sim\mathbb{P}_{\mathcal{C}}, \mathbf{z}\sim Z, \epsilon\sim\mathcal{U}\left(0,1\right)$}
                        \State{$\hat{\mathbf{X}} \gets g_{\Theta}\left(\mathcal{C},\mathbf{z}\right)$}
                        \State{$\tilde{\mathbf{X}} \gets \epsilon{\mathbf{X}} + \left(1-\epsilon\right)\hat{\mathbf{X}}$}
                        \State{$L^i \gets d_{W}(\hat{\mathbf{X}}) - d_{W}\left(\mathbf{X}\right) + \lambda(||\nabla_{\tilde{\mathbf{X}}}d_{W}(\tilde{\mathbf{X}})||_2 - 1)^2$}
                \EndFor
                \State{$W \gets \mathrm{Adam}\left(\nabla_{W}\frac{1}{m}\sum_{i=1}^mL^i,\alpha,\beta_1,\beta_2\right)$}
        \EndFor
        \State{Sample batches $\left\{\mathbf{z}^i\right\}_{i=1}^m\sim Z, \left\{\mathcal{C}^i\right\}_{i=1}^m\sim \mathbb{P}_{\mathcal{C}}$}
        \State{$\{\hat{\mathbf{X}}^i\}_{i=1}^m \gets \left\{g_{\Theta}\left(\mathcal{C}^i,\mathbf{z}^i\right)\right\}_{i=1}^m$ }
        \State{$L \gets \frac{1}{m}\sum_{i=1}^m  - d_{W}(\hat{\mathbf{X}}^i) + \gamma l_{\phi,\mathcal{S}}(\hat{\mathbf{X}}^i)$}
        \State{$\Theta \gets \mathrm{Adam}\left(\nabla_{\Theta}L,\alpha,\beta_1,\beta_2\right)$}
        \Until{$\Theta$ converged}
\end{algorithmic}
\end{algorithm}
While $d_W$ ensures output images look realistic, we use another model to enforce constraint satisfaction.
For this, we introduce the idea of a \emph{constraint critic} that informs the training procedure in a similar manner as $d_W$.
We define the constraint critic loss as the average loss over each constraint after mapping images into the semantic space:
\begin{equation*}
  \label{eq:const_loss}
  \displaystyle l_{\phi,\mathcal{S}}(\hat{\mathbf{X}}, \mathcal{C}) = -\frac{1}{|\mathcal{C}|} \hspace{-0.25em}  \sum_{\left(\mathbf{X}_+,\mathbf{X}_-\right) \in \mathcal{C}} \hspace{-1.25em} p_{\mathcal{S}}(\phi(\hat{\mathbf{X}}),\phi\left(\mathbf{X}_+\right),\phi\left(\mathbf{X}_-)\right)
\end{equation*}
Loss over each constraint $p_{\mathcal{S}}$ is inspired by the loss used in t-Distributed Stochastic Triplet Embedding (STE)~\cite{van2012stochastic}:
\begin{equation*}
  p_{\mathcal{S}} \hspace{-0.1em} \left({a,b,c}\right) = \frac{\left(1+\frac{d_{\mathcal{S}}\left(a,b\right)}{\alpha}\right)^{-\frac{\alpha + 1}{2}}}{\left(1+\frac{d_{\mathcal{S}}\left(a,b\right)}{\alpha}\right)^{-\frac{\alpha + 1}{2}} \hspace{-1.2em} + \left(1+\frac{d_{\mathcal{S}}\left(a,c\right)}{\alpha}\right)^{-\frac{\alpha + 1}{2}}}
\end{equation*}
This loss compares pairs of objects according to a t-Student kernel and is motivated by successes in dimensionality reduction techniques that use heavy tailed similarity kernels~\cite{maaten2008visualizing}.
By minimizing the negation of $p_{\mathcal{S}}$ for each constraint, $\hat{\mathbf{X}}$ is ``pulled'' closer to images $\mathbf{X}_+$ and ``pushed'' farther from images $\mathbf{X}_-$ in $\mathcal{S}$.
As a result, using this loss during training will produce images more likely to satisfy constraints.

We leverage the constraint critic loss in tandem with the discriminator to train the CONGAN generator.
More specifically, our training algorithm is an extension of the Wasserstein GAN~\cite{arjovsky2017wasserstein,gulrajani2017improved}.
We aim to optimize the following:
\begin{equation*}
\min_{\Theta}\max_{W} \hspace{-0.25em} \displaystyle \mathop{\mathbb{E}}_{\mathbf{X} \sim P_{\mathcal{D}}} \hspace{-0.25em} \left[d_{W}\left(\mathbf{X}\right)\right] \hspace{0.1em} - \hspace{-0.5em} \mathop{\mathbb{E}}_{\hat{\mathbf{X}}\sim P_{g}} \hspace{-0.25em} \left[d_{W}(\hat{\mathbf{X}}) - \gamma l_{\phi,\mathcal{S}}(\hat{\mathbf{X}}, \mathcal{C})\right]
\label{eq:WGAN}
\end{equation*}
Here, $P_{\mathcal{D}}$ is a data distribution (i.e. $\mathbf{X}$ is a sample from a training set), $P_{g}$ is the generator distribution (i.e. $\hat{\mathbf{X}} = g_{\Theta}\left(\mathcal{C}, \mathbf{z}\right)$ for a given $\mathbf{z} \sim Z$ and a given $\mathcal{C}$ drawn from a training set of constraint sets).
Finally, $d_W$ is constrained to be 1-Lipschitz. 
This objective is optimized by alternating between updating discriminator parameters $W$ and generator parameters $\Theta$ using stochastic gradient descent, sampling from the training set and generator where necessary.
Intuitively, the discriminator's output can be interpreted as a score of how likely the input is from the data distribution.
When the discriminator updates, it attempts to increase its score for real samples and decrease its score for generated samples.
Conversely, when the generator updates, it attempts to increase the discriminator's score for generated images.
In addition, generator updates decrease the constraint loss by a factor of the hyperparameter $\gamma$.
As a result, generator updates encourage $g_{\Theta}$ to produce images similar to those in the image training set, while also satisfying samples from a constraint training set.
To enforce the 1-Lipschitz constraint on $d_W$ we use the gradient penalty term proposed in~\cite{gulrajani2017improved}.

\begin{figure*}[th]
\begin{center}
   \includegraphics[width=\linewidth]{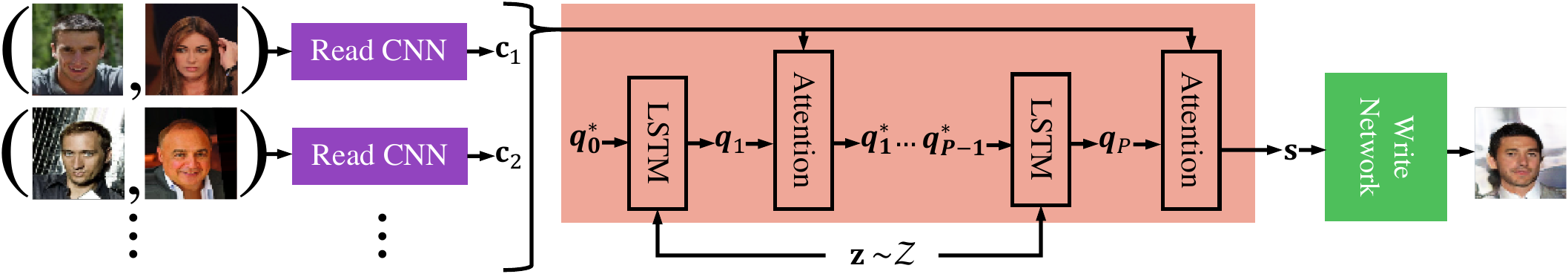}
\end{center}
\caption{The CONGAN generator. Purple is the read network, orange is the process network, and green is the write network}
\label{fig:Generator}
\end{figure*}

The CONGAN training procedure is outlined  in Alg. \ref{alg:CONGAN_train}.
This algorithm is very similar to the WGAN training algorithm (Algorithm 1 in ~\cite{gulrajani2017improved}) with a few key additions.
First, when updating both the discriminator and generator, batches of constraint sets are selected from a training set.
In practice, we use $\phi$ to construct ground truth constraint sets of variable length from images in the image train set, ensuring that our generator is trained on constraint sets that are feasible in $\mathcal{S}$.
Second, the generator update has an additional term: the constraint critic term that encourages constraint satisfaction.

\subsection{A Constrained Generator Network}
While Alg. \ref{alg:CONGAN_train} outlines how to train $g_{\Theta}$, we have yet to formally define $g_{\Theta}$.
In order for $g_{\Theta}$ to accept $\mathcal{C}$ as a \emph{set} it must 1) accept a variable number of constraints, and 2) output the same image regardless of the order in which constraints are given.
For this we leverage the work of~\cite{vinyals2016order} that introduces a neural network framework capable of considering order-invariant inputs, such as sets.
An illustration of the CONGAN generator is depicted in Fig.~\ref{fig:Generator}. 
Our generator has three components: 1) A \emph{read} network used to learn a representation of each constraint 2) a \emph{process} network that combines all constraints in a set into a single set representation, and 3) a \emph{write} network that maps the set representation to an image.
Below we describe each of these components.

The read network puts images within a constraint set through a Convolutional Neural Network (CNN) to extract visual features.
Feature vectors of images from a common constraint pair are concatenated and input to a fully connected layer.
The result is a single vector $\mathbf{c}_i$ for each constraint, which are collectively input to the process network.

The process network consists of a ``processing unit'' that is repeated $p$ times.
Let $\left\{\mathbf{c}_1,...,\mathbf{c}_n\right\}$ be the output of the read network for a size $n$ set of constraints.
For each of the $t$ repetitions of the processing unit, an iteration through an LSTM cell with ``content-based'' attention is performed:
\begin{align}
        \mathbf{q}_t &= LSTM\left(\mathbf{z},\mathbf{q}^{*}_{t-1}\right) \label{eq:LSTM_att}\\
        e_{i,t} &= \mathbf{c}_i \cdot \mathbf{q}_t \label{eq:comb_att}\\
        a_{i,t} &= \frac{\mathrm{exp}\left(e_{i,t}\right)}{\sum_j^n \mathrm{exp}\left(e_{j,t}\right)}\label{eq:softmax_att}\\
        \mathbf{r}_t &= \sum_i^n{a_{i,t}}\mathbf{c}_i\label{eq:sum_att}\\
        \mathbf{q}_t^* &= \left[\mathbf{q}_t,\mathbf{r}_t\right]
\end{align}
First, $\mathbf{z}$ (as ``input'') and the hidden state from  previous repetition are put through LSTM unit.
The resultant hidden state output of the LSTM $\mathbf{q}_t$  is then combined with each $\mathbf{c}_i$ via dot product to create a scalar value $e_{i,t}$ for each constraint.
These are used in a softmax function to obtain scalars $a_{i,t}$, which in turn are used in a weighted sum.
This sum is the key operation that combines the constraints.
Because addition is commutative, the result of \eqref{eq:sum_att}, and thus the output of the processing network, is invariant to the order that the constraints were given.
The result $\mathbf{r}_t$ is concatenated with $\mathbf{q}_t$ and is used as the input in the next processing iteration.
%
After $p$ steps, $\mathbf{q}^{*}_{p}$ is put through a fully connected layer to produce $\mathbf{s}$, which is input to the write network.

One way of interpreting this network is that each processing unit iteration  refines the representation of the constraint set produced by the previous iteration.
The output of the  processing unit has two parts.
First, $\mathbf{r}_t$ is a learned weighted average of the constraints, ideally emphasizing constraints with stronger signal.
Second, $\mathbf{q}_t$ is the output of the LSTM which combines the noise vector and the output from the previous iteration, using various gates to retain certain features while removing others.
These two components are sent back through the processing unit for further rounds of refinement.

%
%
Similar to the generator in the unconditional GAN framework, the write network maps a noise vector to image space.
Motivated by this, we use the transpose convolutions~\cite{dumoulin2016guide,shi2016deconvolution} utilized in Deep Convolutional GANs (DCGANs)~\cite{radford2015unsupervised}.
Transpose convolutions effectively learn an upsampling transformation.
By building a network from transpose convolutional layers, our write network is able to learn how to map from a lower dimensional representation of constraint set to a higher dimensional image.

\begin{figure}[t]
   \centering
   \includegraphics[width=0.9\linewidth]{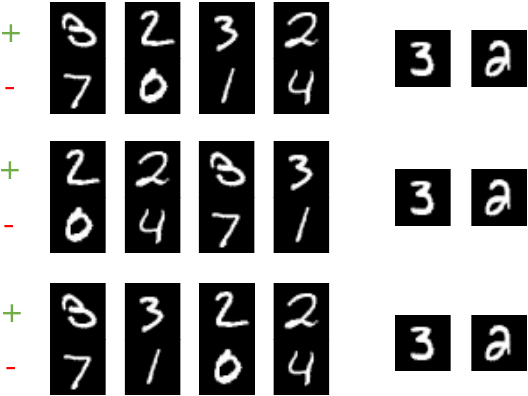}
   \caption{Example illustrating the order invariance property of CONGAN. On the left are relative constraints (top is positive image, bottom is negative) in the order they are input to the CONGAN generator. On the right are the images produced for two different $\mathbf{z}$ vectors. The output remains the same even when the constraints are given in different orders.}
   \label{fig:MNIST_OrderInvar}
\end{figure}

%

\section{Empirical Evaluation}
In order to evaluate CONGAN we aim to show its ability to satisfy constraints while achieving the image quality of similar WGAN models.
Further, we wish to highlight some examples of how a user can interact with a CONGAN generator.
To this end, we perform experiments with three data sets: MNIST~\cite{lecun1998gradient}, CelebA~\cite{yang2015facial}, and Zappos50k~\cite{yu2014fine,yu2017semantic}.

In all experiments, we use the hyperparameters suggested in~\cite{gulrajani2017improved}: ($\lambda = 10, n_{disc} = 5, \alpha = 0.0001, \beta = 0, \beta = 0.9$), follow Algorithm 1 from the same work to train WGAN, and set the batch size $m = 32$.
We seed WGANs with noise vectors $\mathbf{z}$ drawn from a standard normal ($Z = \mathcal{N}\left(0,\mathbf{I}\right)$), and CONGANs with a uniform distribution ($Z = \mathcal{U}\left(-1,1\right)$).
We opt to use the uniform distribution as it allows both inputs into the processing network to be in the same range.
The noise vectors are of size 64 for the MNIST experiments and of size 128 for the CelebA and Zappos50k experiments.
We set $p$ (number of ``processing'' steps) to 5 in both experiments, but have observed that CONGAN is robust to this setting.

In~\cite{arjovsky2017wasserstein} the authors observe that the Wasserstein Distance can be used to determine convergence.
In our experiments, the Wasserstein Distance stopped improving by 100,000 generator update iterations for all models and use that as the iteration limit.
We chose values for $\gamma$ that were able to reduce the t-STE train error significantly while maintaining Wasserstein Distance close to what was achieved by the WGAN.
To strike a good balance  we set $\gamma = 10$ on MNIST, $\gamma = 250$ on CelebA, and $\gamma = 100$ on Zappos50k.

WGAN models were trained on the designated trained sets for MNIST and CelebA.
For Zappos50k, we randomly chose 90\% of the images as the train set, leaving the rest as test.
Similarly, CONGAN model constraint sets $\mathcal{C}$ in the training set of constraint sets are created by first randomly choosing an image of the train set to be a reference image.
Then, anywhere between 1 and 10 pairs of images are randomly chosen to be constraints.
Next, $\phi$ is applied to the reference image and each pair.
The resultant representations in $\mathcal{S}$ are used to determine which elements of the pairs are considered $\mathbf{X}_+$ (positive examples) and $\mathbf{X}_-$ (negative examples) according to~\eqref{eq:const_sat}.
Test sets are constructed similarly.
\begin{figure}[t]
  \centering
  \includegraphics[width=0.9\linewidth]{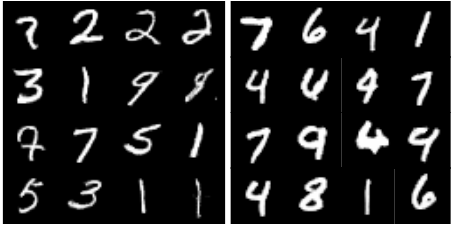}
  \caption{Examples from WGAN (left) and CONGAN (right) generators trained on the MNIST data set.}
  \label{fig:MNIST_Samples}
\end{figure}
\begin{figure*}[th]
  \centering
  \includegraphics[width=.8\linewidth]{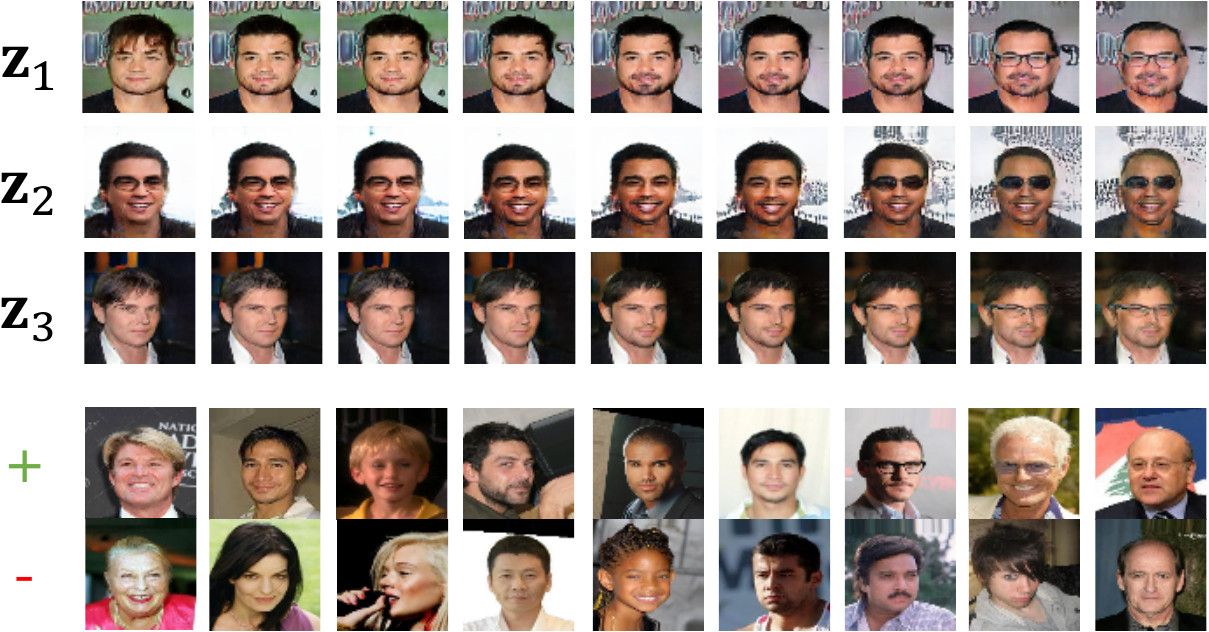}
  \caption{Example of CONGAN generator outputs when trained on the CelebA data set.  The bottom two rows of images are constraints, where the positive and negative images only differ by a single attribute.  The first three constraints differ by only the ``Male'' attribute, the second three by only the ``Beard'' attribute, and the third three by only the ``Eyeglasses'' attribute. The top three rows are images produced from three different seeds when the constraints are provided to the CONGAN generator from left to right.  For example, the third image in the first row is generated when $\mathbf{z}_1$ and the first three constraints are given.}
  \label{fig:CelebAExample1}
\end{figure*}
\begin{figure*}[th]
  \centering
  \includegraphics[width=.8\linewidth]{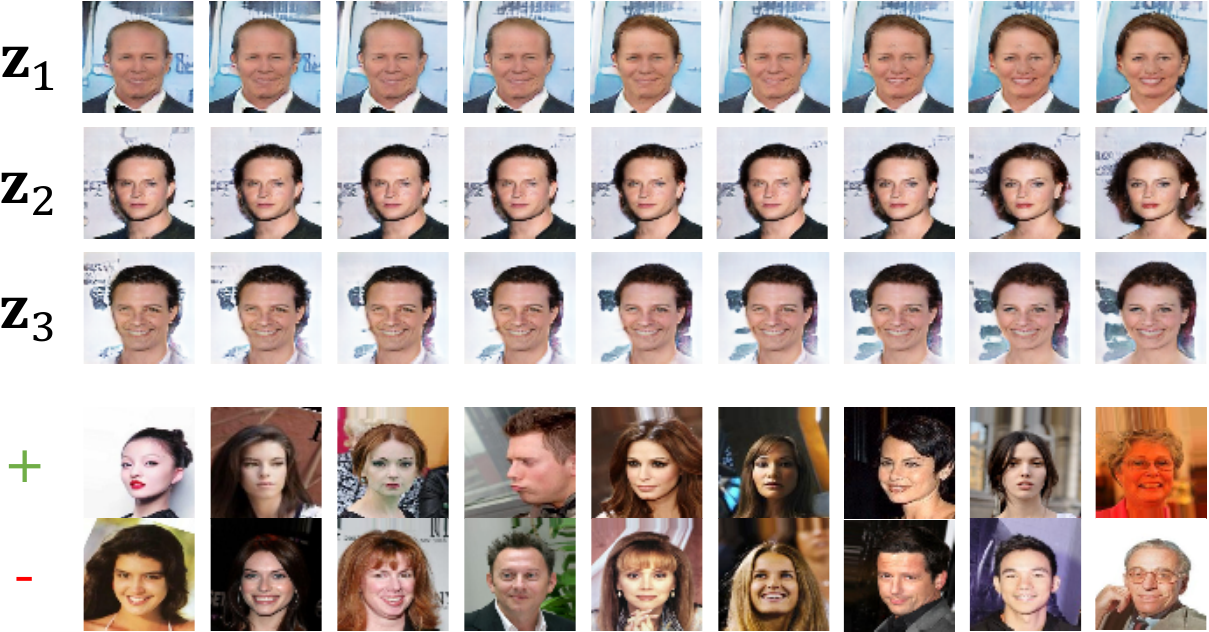}
  \caption{Another example of CONGAN generator outputs when trained on the CelebA data set.  This is the same experiment as in Fig. ~\ref{fig:CelebAExample1}, but with the attributes ``Pale Skin'', ``Brown Hair'', and ``Female'' from left to right.}
  \label{fig:CelebAExample2}
\end{figure*}
%

%
%

The CONGAN network architectures used in these experiments are as follows \footnote{A more rigorous description can be found in the appendix.}.
For MNIST: The discriminator and read networks are five layer CNNs.
The write network is a five layer transpose convolutional network.
For CelebA and Zappos50k: The discriminator and read networks are residual networks~\cite{he2016deep} with four residual CNN blocks.
The write network has four transpose convolutional residual blocks.
To maintain some regularity between models in the interest of fair comparison, we use the same discriminator architectures for both WGAN and CONGAN and use the WGAN generator architecture as the CONGAN write network architecture.
Other than a few special cases, we use rectified linear units as activation functions and perform layer normalization~\cite{ba2016layer}.

\subsection{MNIST}
MNIST is a well known data set containing 28x28 images of hand-written digits.
For preprocessing we zero pad the images to 32x32 and scale them to [-1,1].
For $\phi$ we train a ``mirrored''  autoencoder on the MNIST train set using squared Euclidean loss.
The encoder portion consists of four convolutional layers and a fully connected layer with no activation to a two-dimensional encoding.
We use the encoder as $\phi$.
The decoder has a similar structure but uses transpose convolutions to reverse the mapping.
Simply autoencoding MNIST digits reveals a loose class structure in the embedding space ($\mathcal{S}$ in this experiment).
As such, this experiment shows how class relationships can be retrieved even if $\phi$ does not precisely map to classes.




\begin{table*}
\begin{center}
  \begin{tabular}[th!]{|c|c|c|c|c|c|c|c|c|c|}
    \hline
    \multicolumn{10}{|c|}{\# input constraints} \\
   \multicolumn{1}{|c}{1} & \multicolumn{1}{c}{2} & \multicolumn{1}{c}{3} & \multicolumn{1}{c}{4} & \multicolumn{1}{c}{5} & \multicolumn{1}{c}{6} & \multicolumn{1}{c}{7} & \multicolumn{1}{c}{8} & \multicolumn{1}{c}{9} & \multicolumn{1}{c|}{10}\\ \hline
   0.0931 &  0.0895 & 0.0860 &  0.0831 & 0.0808 & 0.0784 & 0.0775 & 0.0756 & 0.0743 & 0.0733  \\  \hline
\end{tabular}
\end{center}
\caption{Mean constraint satisfaction errors of CONGAN on MNIST constraints per input set size (10 trials).}
\label{tab:MNIST}
\end{table*}
%
%
\begin{table*}
\begin{tabular}[th!]{l|r|r|r|r|r|r|r|r|r|r|r|}
   \cline{2-12}      
   &   & \multicolumn{10}{c|}{CONGAN (\# input constraints)} \\ 
   & \multicolumn{1}{c|}{WGAN} &  \multicolumn{1}{c}{1} & \multicolumn{1}{c}{2} & \multicolumn{1}{c}{3} & \multicolumn{1}{c}{4} & \multicolumn{1}{c}{5} & \multicolumn{1}{c}{6} & \multicolumn{1}{c}{7} & \multicolumn{1}{c}{8} & \multicolumn{1}{c}{9} & \multicolumn{1}{c|}{10} \\ \hline 
   \multicolumn{1}{|l|}{-WGAN}  & 20.31& 18.32 & 18.90 &  19.34 & 19.64 & 19.82 & 19.93 & 20.00 & 19.99 & 19.96 & 19.90  \\ 
   \multicolumn{1}{|l|}{-CONGAN}  & 481.12 & 479.27 & 480.08 & 480.74 & 481.29 & 481.74 & 482.07 & 482.36 & 482.57 & 482.71 & 482.80 \\ \Xhline{2\arrayrulewidth} 
   \multicolumn{1}{|l|}{MCSE} & & 0.0885 &  0.1047 & 0.1154 &  0.1202 &  0.1257 &  0.1279 & 0.1296 &  0.1307 &  0.1318 &   0.1325 \\ \hline
   \multicolumn{12}{c}{ }
\end{tabular}

\begin{tabular}[th!]{l|r|r|r|r|r|r|r|r|r|r|r|}
   \cline{2-12}      
   &   & \multicolumn{10}{c|}{CONGAN (\# input constraints)} \\ 
   & \multicolumn{1}{c|}{WGAN} &  \multicolumn{1}{c}{1} & \multicolumn{1}{c}{2} & \multicolumn{1}{c}{3} & \multicolumn{1}{c}{4} & \multicolumn{1}{c}{5} & \multicolumn{1}{c}{6} & \multicolumn{1}{c}{7} & \multicolumn{1}{c}{8} & \multicolumn{1}{c}{9} & \multicolumn{1}{c|}{10} \\ \hline 
   \multicolumn{1}{|l|}{WGAN} & 60.31 & 44.85 & 46.03 & 46.99 & 47.45 & 47.30 & 46.76 & 45.84 & 44.86 & 43.81 & 42.78 \\ 
   \multicolumn{1}{|l|}{CONGAN}  & 5.43 & -27.04 & -18.64 & -11.39 & -5.53 & -1.37 & 1.09 &  2.06 &  1.69 & 0.10 & -2.41\\ \Xhline{2\arrayrulewidth} 
   \multicolumn{1}{|l|}{MCSE} & & 0.0950 & 0.0967 & 0.0974 & 0.1001 & 0.1009 & 0.1019 & 0.1052 & 0.1065 &  0.1065 & 0.1066 \\ \hline
\end{tabular}
\caption{Evaluation results on CelebA (top table) and Zappos50K (bottom table) data sets (10 trials). Rows 1-2 of each table: Mean discriminator scores for WGAN and CONGAN discriminators at convergence on WGAN and CONGAN generators (negative scores for CelebA).  Row 3 of each table: Mean constraint satisfaction error of CONGAN models per input set size.}
\label{tab:CelebA}
\end{table*}

%




We seek to evaluate the CONGAN's ability to satisfy given constraints.
To this end, we constructed ten different test sets, each containing constraint sets of a fixed size.
For example, each constraint set in the ``2'' test set has two constraints.
We call an evaluation over a different test set an ``experiment''.
In each experiment, we performed ten different trials where the generator was given different noise vectors per constraint set.
With these experiments we can observe the effect constraint set size has on the generator.

\textbf{Results:} Table \ref{tab:MNIST} shows the mean constraint satisfaction error (i.e one minus the prevalence of \eqref{eq:const_sat})
of the CONGAN generator for each MNIST experiment.
Overall, it was able to satisfy over 90\% of given constraints.
Note that the generator performs slightly better when more constraints are given.
This is somewhat counter-intuitive.
We believe that in this case the generator is using constraints to determine what class of digit to  produce.
If given few constraints, it is more difficult for the generator to determine the class of the output.
Figures \ref{fig:MNIST_OrderInvar} and \ref{fig:MNIST_Samples} show example outputs of CONGAN when trained on MNIST: One showing the order invariance property of CONGAN and the other showing CONGAN generated images next to ones produced by a similar WGAN.

\subsection{CelebA}
The CelebA data set contains 202,599  color images of celebrity faces.
For our experiments, we resize each image to 64x64 and scale to [-1,1].
Associated with each image are 40 binary attributes ranging from ``Blond Hair'' to ``Smiling''.
We chose twelve of these attributes to be $\mathcal{S}$.
More specifically, an image's representation in $\mathcal{S}$ is a binary vector of attributes, which differs from the MNIST experiment.
In the previous experiment, $\mathcal{S}$  was both lower dimensional and continuous.
As such, this experiment will evaluate CONGAN's ability to adapt to different semantic spaces.

For $\phi$ we construct a simple multi-task CNN (MCNN)~\cite{hand2017attributes} \footnote{Details and an evaluation of the MCNN can be found in the appendix.} that consists of one base network and multiple specialized networks, trained end-to-end.
The base network accepts the image as input and extracts features for detecting all attributes.
The specialized networks split from the base network and learn to detect to their predetermined subset.
Our $\phi$ base network consists of two convolutional layers.
The specialized networks (one for each of twelve attributes) consists of three convolutional layers followed by a fully connected layer that maps to a scalar attribute identifier.




For this experiment we sought to more objectively compare the WGAN generated images with those produced by CONGAN.
To this end we first train a WGAN on the CelebA train set.
Then, we initialize the CONGAN write network and discriminator to the trained WGAN generator and discriminator, respectively, before training the CONGAN generator.
By doing this, we can observe how image quality is affected by adding the CONGAN components to a WGAN.

\textbf{Results:} Rows one and two of Table~\ref{tab:CelebA} (top table) show the mean negative discriminator scores for both the WGAN and CONGAN generators against the WGAN and CONGAN discriminators at convergence over ten trials.
We can see that for both discriminators, WGAN generated images are scored very similarly to those generated by CONGAN.
This is especially true when considering the standard deviation for the WGAN generator against the WGAN and CONGAN discriminators is 8.75 and 16.22, respectively, and slightly higher on both for the CONGAN generator.
We believe this result shows evidence that adding the CONGAN framework to the WGAN training did not drastically alter image quality.


%
\begin{figure}[t]
  \centering
  \includegraphics[width=0.9\linewidth]{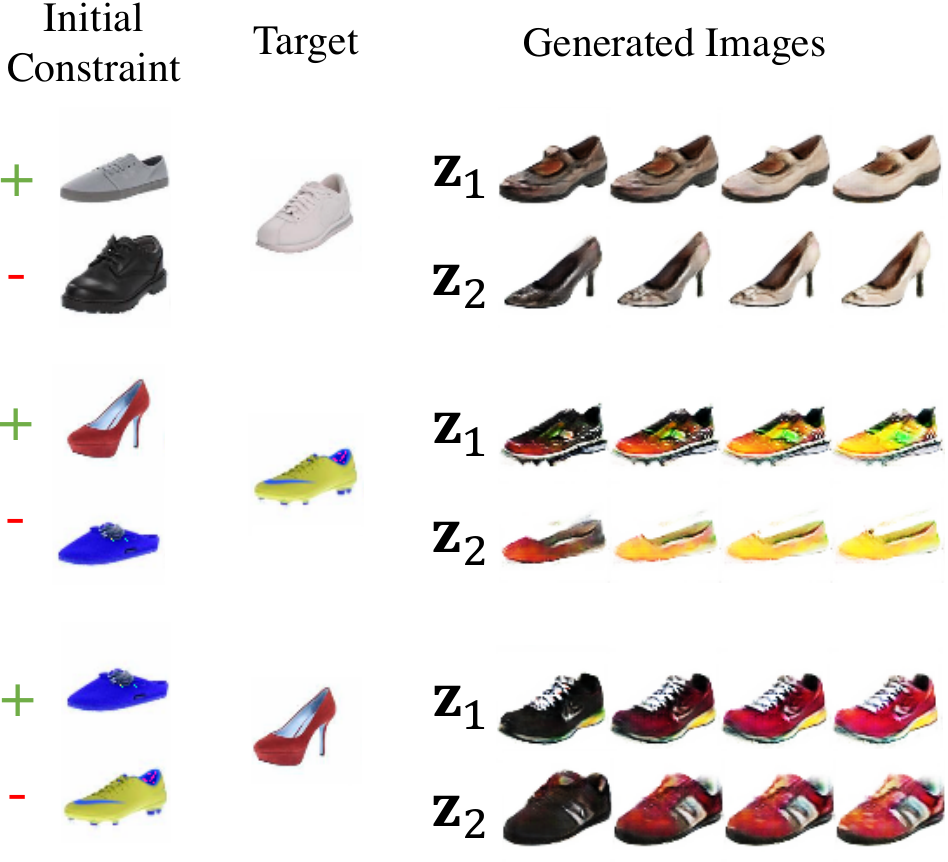}
  \caption{Three sets of two examples from the CONGAN generator trained on the Zappos data set. The generator was first provided the initial constraint of the left, generating the first (left-most) image in the generated images column.  To generate each of the next three images, the generator was fed a constraint where the positive image was the target image, and the negative image was the previously generated image.}
  \label{fig:Zappos_Samples}
\end{figure}

The last row of Table~\ref{tab:CelebA} shows the mean constraint satisfaction error on the test set for each experiment.
Here, the CONGAN generator is able to satisfy around 87\% or more of the constraints.
Figures~\ref{fig:CelebAExample1} and~\ref{fig:CelebAExample2} show images generated by CONGAN.
As constraints are provided, the image produced from different seeds take on the attributes indicated by the constraints.
In Fig. ~\ref{fig:CelebAExample1}, the first three constraints indicate the ``Male'' attribute, the next three indicate ``Beard'', and the last ``Eyeglasses''.
In Fig. ~\ref{fig:CelebAExample2}, ``Pale Skin'', ``Brown Hair'', and ``Female'' are indicated.
These examples show that a user can iteratively refine the images to have desired characteristics, and still be given a variety of realistic, novel images.

\subsection{Zappos50K}
The Zappos50K data set contains 50,025 color images of shoes.
We resize each image to 64x64 and scale to [-1,1].
For this experiment, we chose $\mathcal{S}$ to be a color space.
To accomplish this, we computed a 64 bin color histogram over each image and trained a nine-layer CNN to embed the images in 2-dimensions using a triplet network~\cite{hoffer2015deep} \footnote{A visualization of this embedding can be found in the appendix.}, and used this as $\phi$.
We opted to use the T-STE loss in the objective as it produced a clear separation of colors.

There is inherent bias in the Zappos50K data set when it comes to color, as most shoes tend to be black, brown, or white.
This poses a challenge in that if constraint sets used for training  are formed by uniformly sampling over the train set, the model will tend to favor few colors, making it difficult to guide generation to other colors.
To combat this, we constructed constraints to include a more uniform sampling over colors.
When constructing the train set of constraint sets, with probability 0.5 we uniformly sampled over training images as in the other experiments.
When not sampling uniformly,
we focused on a single color bin by first selecting a bin and choosing all positive images in the constraint set to be images where the highest histogram value corresponded to that bin (e.g. all positive examples would be ``light blue'').
Negative examples would be chose uniformly from the other bins.
We found this allowed the CONGAN generator to more easily learn to produce a variety of colors.

\textbf{Results:} Table~\ref{tab:CelebA} (bottom table) shows the discriminator scores and mean constraint satisfaction errors for each Zappos50K experiment.
Here, the CONGAN generator produced lower scores than the WGAN for both discriminators, though within one standard deviation.
We believe this is due to training the generator to produce a wider variety of colors.
If training data contains many brown, black, and white shoes, then training the generator to produce blue, red and yellow shoes will force it to produce images that differ than those provided to the discriminator.
Nevertheless, we believe that image quality was only slightly degraded as a result.

Figure~\ref{fig:Zappos_Samples} shows examples of the images produced by the CONGAN generator.
Here, we wanted to test the use case of providing single images, instead of pair-wise constraints, to guide the generator to a result.
An initial constraint is provided to produce a starting images.
After that, a single target image is used repeatedly as the positive example to generate shoes more similarly colored to the target.




%

\section{Conclusion and Future Work}
In this work, we introduce a Generative Adversarial Network framework that is able to generate imagery guided by iterative human feedback.
Our model relies on two novel components.
First, we develop a generator, based on recent work in memory networks, that maps variable-sized sets of constraints to image space using order-invariant operations.
Second, this generator is informed during training by a critic that determines whether generated imagery satisfies given constraints.
The result is a generator that can can be guided interactively by humans through relative constraints.
Empirically our model is able to generate images that are of comparable quality to those produced by similar GAN models, while satisfying a up to 90\% of given constraints.

There are multiple avenues of future work that we believe are worthy of further study.
First, it may not be feasible for users of CONGAN to search through large image databases to find the exact constraints they desire.
We will apply pair-wise active ranking techniques~\cite{jamieson2011active} to suggest constraint queries in order to quickly constrain the semantic space without requiring users to search through images themselves.
Second,
we will investigate the output of the process network more closely seeing if constraint representations have properties that match intuition about how sets of constraints are classically reasoned about, similar to word embeddings~\cite{mikolov2013distributed}.

\textbf{Acknowledgments:} This work was supported by the AFOSR Science of Information, Computation, Learning, and Fusion program lead by Dr. Doug Riecken. Eric would like to thank Davis Gilton (UW-M) and Timothy Van Slyke (NEU) for early exploratory experiments that made this work possible.  Eric would also like to thank Ritwik Gupta (SEI/CMU) for reviewing a draft of the paper.  Finally, Eric would like to thank his colleagues at AFRL/RI: Dr. Lee Seversky, Dr. Walter Bennette, Dr. Matthew Klawonn, and Dylan Elliot for insightful feedback as this work progressed.

\clearpage

{\small
\bibliographystyle{ieee}
\bibliography{congan}
}

\appendix

\section{Formal Definition of LSTM Component}

Equation (4) is a standard LSTM cell:
\begin{align*}
        LSTM\left(\mathbf{z},\mathbf{q}^{*}_{t-1}\right) &= \mathbf{o}_t * \mathrm{tanh}\left(\mathbf{h}_t\right)\mathrm{, where}\\
        \mathbf{o}_t &= \sigma\left(\mathbf{w_o}\cdot\left[\mathbf{q}^{*}_{t-1},\mathbf{z}\right] + \mathbf{b_o}\right)\\
        \mathbf{h}_t &= \mathbf{f}_t*\mathbf{h}_{t-1}+\mathbf{i}_t*\tilde{\mathbf{h}}_t\\
        \mathbf{f}_t &= \sigma\left(\mathbf{w_f}\cdot\left[\mathbf{q}^{*}_{t-1},\mathbf{z}\right] + \mathbf{b_f}\right)\\
        \mathbf{i}_t &= \sigma\left(\mathbf{w_i}\cdot\left[\mathbf{q}^{*}_{t-1},\mathbf{z}\right] + \mathbf{b_i}\right)\\
        \tilde{\mathbf{h}}_t &= \mathrm{tanh}\left(\mathbf{w}_{\tilde{\mathbf{h}}}\cdot\left[\mathbf{q}^{*}_{t-1},\mathbf{z}\right] + \mathbf{b}_{\tilde{\mathbf{h}}}\right)
\end{align*}
Here, $\mathbf{z}$ is used as what is commonly referred to as ``input'' to the LSTM, $\mathbf{q}^{*}_{t-1}$ is commonly called the ``hidden state'' of the previous iteration, and $LSTM$ returns the hidden state of the current iteration.

\section{Neural Network Architectures used in Experiments}
In this section, we outline the neural network architecture used in all experiments in the main paper, layer by layer.
Rows of the network in descending order (top to bottom) indicate layers from input to output.
The following naming conventions are used throughout.
``Conv'' indicates a convolutional layer, ``FC'' indicates a fully connected layer, and ``TConv'' indicates a transpose convolutional layer.
The column labeled ``Ker'' indicates the kernel size, ``Str'' indicates stride, and ``Act'' indicates the activation function used.
Columns labeled ``In'' and ``Out'' indicate the shape of the input to the layer and shape of the output of the layer.

\subsection{MNIST Experiments}
Below you will find architecture descriptions for the networks used in the MNIST experiments. Note that after each two convolutional or transpose convolutional layers in all networks, layer normalization is used.
\begin{table}[H]
\begin{center}
\begin{tabular}[t]{|l|l|l|l|l|l|}
   \hline
   \multicolumn{6}{|c|}{$\phi$ \textbf{Network (Encoder)}} \\ \hline
   \textbf{Layer} & \textbf{In} & \textbf{Ker} & \textbf{Str} & \textbf{Act} & \textbf{Out} \\ \hline
   Conv & 32x32x1 & 3x3 & 1 & ReLU & 32x32x4 \\ \hline
   Conv & 32x32x4 & 3x3 & 2 & ReLU & 16x16x8 \\ \hline
   Conv & 16x16x8 & 3x3 & 2 & ReLU & 8x8x16 \\ \hline
   Conv & 8x8x16 & 3x3 & 2 & ReLU & 4x4x32 \\ \hline
   Conv & 4x4x32 & 3x3 & 2 & ReLU & 2x2x64 \\ \hline
   FC & 2x2x64 &  &  & None  & 2 \\ \hline 
\end{tabular}
\end{center}
\vspace{-2em}
\end{table}
\begin{table}[H]
\begin{center}
\begin{tabular}[t]{|l|l|l|l|l|l|}
   \hline
   \multicolumn{6}{|c|}{$\phi$ \textbf{Network (Decoder)}} \\ \hline
   \textbf{Layer} & \textbf{In} & \textbf{Ker} & \textbf{Str} & \textbf{Act} & \textbf{Out} \\ \hline
   FC & 2 &  &  &  None & 2x2x64 \\ \hline
   TConv & 2x2x64 & 3x3 & 2 & ReLU & 4x4x32 \\ \hline
   TConv & 4x4x32 & 3x3 & 2 & ReLU & 8x8x16 \\ \hline
   TConv & 8x8x16 & 3x3 & 2 & ReLU & 16x16x8 \\ \hline
   TConv & 16x16x8 & 3x3 & 2 & ReLU & 32x32x4 \\ \hline
   Conv & 32x32x4 & 3x3 & 1 & tanh & 32x32x1 \\ \hline
\end{tabular}
\end{center}
\vspace{-2em}
\end{table}
\begin{table}[H]
\begin{center}
\begin{tabular}[t]{|l|l|l|l|l|l|}
   \hline
   \multicolumn{6}{|c|}{\textbf{Discriminator Network (WGAN and CONGAN)}} \\ \hline
   \textbf{Layer} & \textbf{In} & \textbf{Ker} & \textbf{Str} & \textbf{Act} & \textbf{Out} \\ \hline
   Conv & 32x32x1 & 3x3 & 1 & ReLU & 32x32x64 \\ \hline
   Conv & 32x32x64 & 3x3 & 2 & ReLU & 16x16x128 \\ \hline
   Conv & 16x16x128 & 3x3 & 2 & ReLU & 8x8x256 \\ \hline
   Conv & 8x8x256 & 3x3 & 2 & ReLU & 4x4x512 \\ \hline
   FC & 4x4x512 &  &  & None  & 1 \\ \hline 
\end{tabular}
\end{center}
\vspace{-2em}
\end{table}
\begin{table}[H]
\begin{center}
\begin{tabular}[t]{|l|l|l|l|l|l|}
   \hline
   \multicolumn{6}{|c|}{\textbf{Read CNN}} \\ \hline
   \textbf{Layer} & \textbf{In} & \textbf{Ker} & \textbf{Str} & \textbf{Act} & \textbf{Out} \\ \hline
   Conv & 32x32x1 & 5x5 & 1 & ReLU & 32x32x2 \\ \hline
   Conv & 32x32x2 & 5x5 & 2 & ReLU & 16x16x4 \\ \hline
   Conv & 16x16x4 & 5x5 & 2 & ReLU & 8x8x8 \\ \hline
   Conv & 8x8x8 & 5x5 & 2 & ReLU & 4x4x16 \\ \hline
   Conv & 4x4x16 & 5x5 & 2 & ReLU & 2x2x32 \\ \hline
   FC & 2x2x32 &  &  & tanh & 64 \\ \hline 
\end{tabular}
\end{center}
\vspace{-2em}
\end{table}
\begin{table}[H]
\begin{center}
\begin{tabular}[t]{|l|l|l|l|l|l|}
   \hline
   \multicolumn{6}{|c|}{\textbf{CONGAN Write Network/WGAN Generator}} \\ \hline
   \textbf{Layer} & \textbf{In} & \textbf{Ker} & \textbf{Str} & \textbf{Act} & \textbf{Out} \\ \hline
   FC & 64 &  &  & None & 4x4x512 \\ \hline 
   TConv & 4x4x512 & 3x3 & 2 & ReLU & 8x8x256 \\ \hline
   Conv & 8x8x256 & 3x3 & 1 & ReLU & 8x8x256 \\ \hline
   TConv & 8x8x256 & 3x3 & 2 & ReLU & 16x16x128 \\ \hline
   Conv & 16x16x128 & 3x3 & 1 & ReLU & 16x16x128 \\ \hline
   TConv & 16x16x128 & 3x3 & 2 & ReLU & 32x32x64 \\ \hline
   Conv & 32x32x64 & 3x3 & 1 & tanh & 32x32x1 \\ \hline
\end{tabular}
\end{center}
\end{table}

\subsection{CelebA and Zappos50K Experiments}
In this section, we first describe all network architectures used in both the CelebA and Zappos50K experiments.
Then we outline the $\phi$ networks used for each.
Here, ``Norm'' indicates layer norm, ``ReLU'' indicates the application of a rectified linear unit.
The ``ID'' column is used to identify which layers are used in subsequent operations in the residual block.
For the residual blocks, the ``In'' column is either used to indicate the size of the input or the IDs of the layers used as input.
The ``Add'' layers are simply the addition of the two layers identified in the ``In'' column with the first ID multiplied by 0.3 before the addition.
The ``RB$\uparrow$'' layer is a residual block up and ``RB$\downarrow$'' is a residual block down.

\begin{table}[H]
\begin{center}
\begin{tabular}[t]{|l|l|l|l|l|l|}
   \hline
   \multicolumn{6}{|c|}{\textbf{Discriminator Network}} \\ \hline
   \textbf{Layer} & \textbf{In} & \textbf{Ker} & \textbf{Str} & \textbf{Act} & \textbf{Out} \\ \hline
   Conv & 64x64x3 & 3x3 & 1 & ReLU & 64x64x64 \\ \hline
   RB$\downarrow$ & 64x64x64 & & & & 32x32x128 \\ \hline
   RB$\downarrow$ & 32x32x128 & & & & 16x16x256 \\ \hline
   RB$\downarrow$ & 16x16x256 & & & & 8x8x512 \\ \hline
   RB$\downarrow$ & 8x8x512 & & & & 4x4x512 \\ \hline
   FC & 4x4x512 &  &  & None & 1  \\ \hline       
\end{tabular}
\end{center}
\vspace{-2em}
\end{table}
\begin{table}[H]
\begin{center}
\begin{tabular}[t]{|l|l|l|l|l|l|}
   \hline
   \multicolumn{6}{|c|}{\textbf{Read CNN}} \\ \hline
   \textbf{Layer} & \textbf{In} & \textbf{Ker} & \textbf{Str} & \textbf{Act} & \textbf{Out} \\ \hline
   Conv & 64x64x3 & 3x3 & 1 & ReLU & 64x64x8 \\ \hline
   RB$\downarrow$ & 64x64x8 & & & & 32x32x16 \\ \hline
   RB$\downarrow$ & 32x32x16 & & & & 16x16x32 \\ \hline
   RB$\downarrow$ & 16x16x32 & & & & 8x8x32 \\ \hline
   FC & 8x8x32 &  &  & tanh & 1  \\ \hline       
\end{tabular}
\end{center}
\vspace{-2em}
\end{table}
\begin{table}[H]
\begin{center}
\begin{tabular}[t]{|l|l|l|l|l|l|}
   \hline
   \multicolumn{6}{|c|}{\textbf{CONGAN Write Network/WGAN Generator}} \\ \hline
   \textbf{Layer} & \textbf{In} & \textbf{Ker} & \textbf{Str} & \textbf{Act} & \textbf{Out} \\ \hline
   FC & 128 & &  & ReLU & 4x4x512 \\ \hline
   RB$\uparrow$ & 4x4x512 & & & & 8x8x512 \\ \hline
   RB$\uparrow$ & 8x8x512 & & & & 16x16x256 \\ \hline
   RB$\uparrow$ & 16x16x256 & & & & 32x32x128 \\ \hline
   RB$\uparrow$ & 32x32x128 &  & & & 64x64x64 \\ \hline
   Conv & 64x64x64 & 3x3 & 1 & tanh & 64x64x3  \\ \hline       
\end{tabular}
\end{center}
\vspace{-2em}
\end{table}
\begin{table}[H]
\begin{center}
\begin{tabular}[t]{|l|l|l|l|l|l|l|}
   \hline
   \multicolumn{7}{|c|}{\textbf{Residual Block (Down)}} \\ \hline
   \textbf{ID} & \textbf{Layer} & \textbf{In} & \textbf{Ker} & \textbf{Str} & \textbf{Act} & \textbf{Out} \\ \hline
   1 & Conv & $a$x$b$x$c$ & 5x5 & 2 & None & $\frac{a}{2}$x$\frac{b}{2}$x$d$ \\ \hline
   2 & Conv & $a$x$b$x$c$ & 5x5 & 1 & None & $a$x$b$x$c$ \\ \hline
   3 & Norm & (2) & & & & \\ \hline
   4 & ReLU & (3) & & & & \\ \hline
   5 & Conv & (4) & 5x5 & 2 & None & $\frac{a}{2}$x$\frac{b}{2}$x$d$ \\ \hline
   6 & Add & (5), (1) & & & & \\ \hline
   7 & Norm & (6) & & & & \\ \hline
   8 & ReLU & (7) & & & & \\ \hline
 \end{tabular}
\end{center}
\vspace{-2em}
\end{table}
\begin{table}[H]
\begin{center}
\begin{tabular}[t]{|l|l|l|l|l|l|l|}
   \hline
   \multicolumn{7}{|c|}{\textbf{Residual Block (Up)}} \\ \hline
   \textbf{ID} & \textbf{Layer} & \textbf{In} & \textbf{Ker} & \textbf{Str} & \textbf{Act} & \textbf{Out} \\ \hline
   1 & TConv & $a$x$b$x$c$ & 5x5 & 2 & None & $(2*a)$x \\ 
   & & & & & & $(2*b)$x \\
   & & & & & & $d$ \\ \hline
   2 & Conv & $a$x$b$x$c$ & 5x5 & 1 & None & $a$x$b$x$c$ \\ \hline
   3 & Norm & (2) & & & & \\ \hline
   4 & ReLU & (3) & & & & \\ \hline
   5 & TConv & (4) & 5x5 & 2 & None & $(2*a)$x \\
   & & & & & & $(2*b)$x \\
   & & & & & & $d$ \\ \hline
   6 & Add & (5), (1) & & & & \\ \hline
   7 & Norm & (6) & & & & \\ \hline
   8 & ReLU & (7) & & & & \\ \hline
 \end{tabular}
\end{center}
\end{table}

\subsection{Celeba $\phi$ MCNN}
The MCNN we developed for the $\phi$ network in our CelebA experiments takes an image, and puts it through a ``base'' network.  Then the output of the base network is input to twelve``specialized'' networks to predict the presence or absence of each of the twelve attributes we used in our experiment. Each of these architectures are outlined below.
\begin{table}[H]
\begin{center}
\begin{tabular}[t]{|l|l|l|l|l|l|}
   \hline
   \multicolumn{6}{|c|}{$\phi$ \textbf{MCNN Network (Base)}} \\ \hline
   \textbf{Layer} & \textbf{In} & \textbf{Ker} & \textbf{Str} & \textbf{Act} & \textbf{Out} \\ \hline
   Conv & 64x64x3 & 7x7 & 2 & ReLU & 32x32x64 \\ \hline
   Conv & 32x32x64 & 5x5 & 2 & ReLU & 16x16x128 \\ \hline      
   Norm & & & & & \\ \hline
\end{tabular}
\end{center}
\vspace{-2em}
\end{table}
\begin{table}[H]
\begin{center}
\begin{tabular}[t]{|l|l|l|l|l|l|}
   \hline
   \multicolumn{6}{|c|}{$\phi$ \textbf{MCNN Network (Specialized)}} \\ \hline
   \textbf{Layer} & \textbf{In} & \textbf{Ker} & \textbf{Str} & \textbf{Act} & \textbf{Out} \\ \hline
   Conv & 16x16x128 & 3x3 & 2 & ReLU & 8x8x256 \\ \hline
   Conv & 8x8x256 & 3x3 & 2 & ReLU & 4x4x512 \\ \hline
   Norm & & & & & \\ \hline
   Conv & 4x4x512 & 3x3 & 2 & ReLU & 2x2x1024 \\ \hline
   FC & 2x2x1024 &  &  & sigm & 1 \\ \hline       
\end{tabular}
\end{center}
\end{table}
\begin{figure}[t]
\begin{center}
   \includegraphics[width=\linewidth]{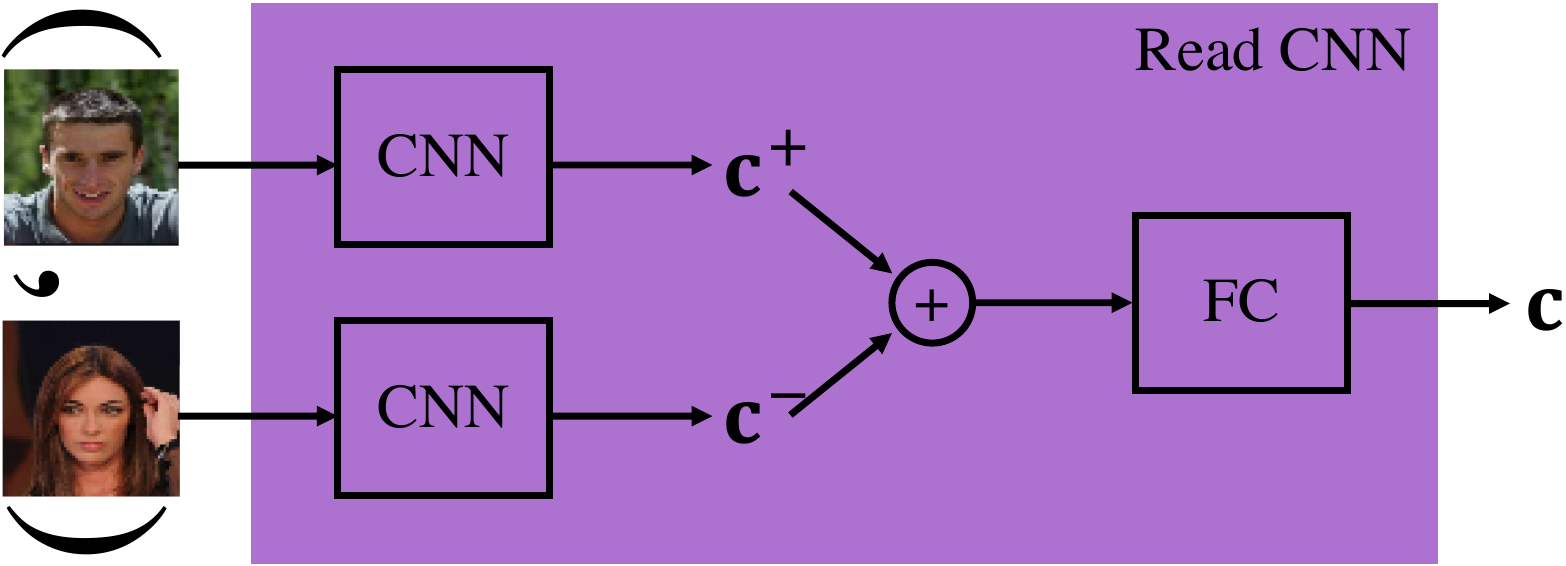}
\end{center}
\caption{The read network to map a constraint to a vector.}
\label{fig:ReadNet}
\end{figure}
\begin{figure}[t]
\begin{center}
   \includegraphics[width=.94\linewidth]{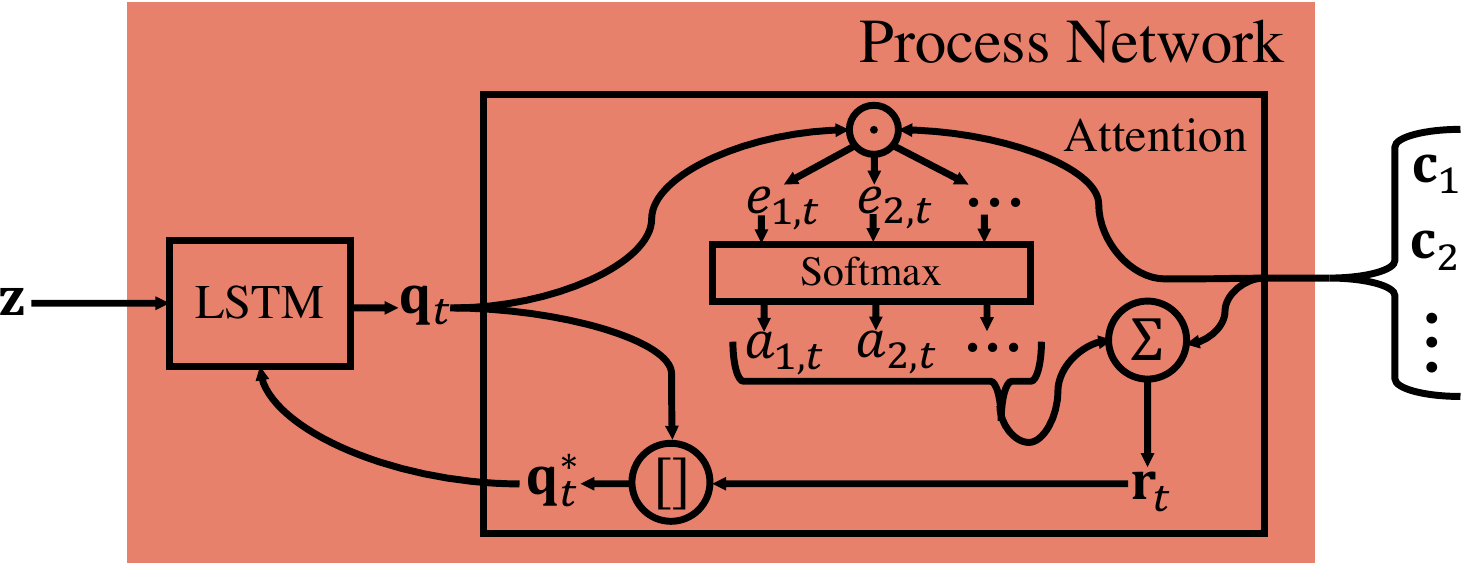}
\end{center}
\caption{Illustration of the $t$th iteration of the process network, beginning with the LSTM unit and ending with $q^*_t$.}
\label{fig:ProcessNet}
\end{figure}

\subsection{Zappos50K $\phi$ Triplet Network}
A triplet network takes three images and puts them through the same network resulting in an $n$ dimensional embedding for which standard triplet losses can be applied.  Below describes the network we used in our Zappos50K experiments. Note that after each two convolutional layers, layer normalization is applied.
\begin{table}[H]
\begin{center}
\begin{tabular}[t]{|l|l|l|l|l|l|}
   \hline
   \multicolumn{6}{|c|}{$\phi$ \textbf{Triplet Network}} \\ \hline
   \textbf{Layer} & \textbf{In} & \textbf{Ker} & \textbf{Str} & \textbf{Act} & \textbf{Out} \\ \hline
   Conv & 64x64x3 & 5x5 & 1 & ReLU & 64x64x8 \\ \hline
   Conv & 64x64x8 & 5x5 & 2 & ReLU & 32x32x8 \\ \hline
   Conv & 32x32x8 & 5x5 & 1 & ReLU & 32x32x16 \\ \hline
   Conv & 32x32x16 & 5x5 & 2 & ReLU & 16x16x16 \\ \hline
   Conv & 16x16x16 & 5x5 & 1 & ReLU & 16x16x32 \\ \hline
   Conv & 16x16x32 & 5x5 & 2 & ReLU & 8x8x32 \\ \hline
   Conv & 8x8x32 & 5x5 & 1 & ReLU & 8x8x64 \\ \hline
   Conv & 8x8x64 & 5x5 & 2 & ReLU & 4x4x64 \\ \hline
   FC & 4x4x64 &  &  & None & 2 \\ \hline       
\end{tabular}
\end{center}
\end{table}

\section{CelebA $\phi$ MCNN Training Details and Performance}
For training the $\phi$ MCNN used in the CelebA data experiments, we chose twelve attributes for the network to predict.
We used the Adam optimization method with default parameters, a batch size of 32, and trained the model for 100,000 iterations.
The test accuracy of the network for the twelve attributes is shown in the table below.
We note that these results are slightly worse than those reported in the original paper, but sufficient for the CONGAN generator to learn how to manipulate images.
Performance can be increased by employing the ``aux'' method described in the original MCNN paper, and by designing the architecture to be take advantage of groups of common attributes.

\begin{table}[H]
\begin{center}
\begin{tabular}[H]{|l|r|}
\hline
\multicolumn{1}{|c|}{\textbf{Attribute}} & \multicolumn{1}{|c|}{\textbf{Accuracy}} \\ \hline
Bald &  0.9836\\ \hline
Black Hair & 0.8870 \\ \hline
Blond Hair & 0.9414 \\ \hline
Brown Hair & 0.8242 \\ \hline
Eyeglasses & 0.9901 \\ \hline
Goatee & 0.9531 \\ \hline
Gray Hair & 0.9709 \\ \hline
Male & 0.9760 \\ \hline
Mustache & 0.9557 \\ \hline
No Beard & 0.9360 \\ \hline
Pale Skin & 0.9601 \\ \hline
Wearing Hat &  0.9832 \\ \hline
\end{tabular}
\label{tab:MCNN}
\end{center}
\end{table}

\section{Zappos50K $\phi$ Triplet Network Training Details and Performance}
We formed the training set for the triplet network by first taking each image in the Zappos50K train set, and placed it into one of the 64 color histogram bins according their highest histogram value.  To form each triplet $\left(A,B,C\right)$ (``$A$ is more similar to $C$ than $C$''), we iterated over each bin $j$, selecting images $A$ and $B$ randomly from $j$, and image $C$ randomly from another bin.  We iterated over each bin 5000 times creating 320,000 triplets for training.  We did a similar process for the test set, but with 1000 ``passes'' over each bin,, making a test set of 64,000 triplets.

\begin{sidewaysfigure*}[t]
\begin{center}
   \includegraphics[width=\linewidth]{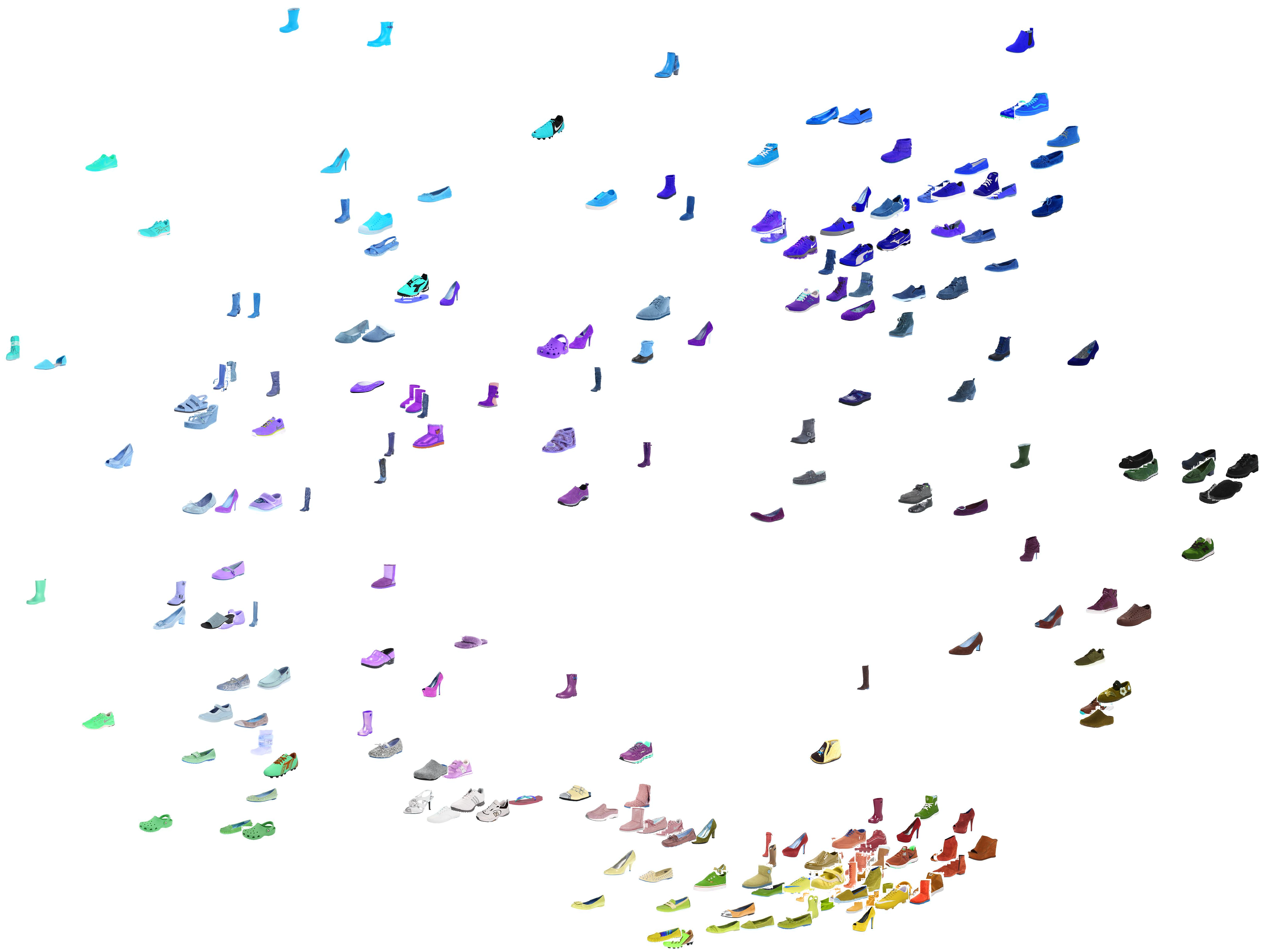}
\end{center}
\caption{Samples from the Zappos50K data set embedded using the $\phi$ triplet network.}
\label{fig:ZapposEmbedding}
\end{sidewaysfigure*}

We trained the network using default Adam optimization parameters and a batch size of 128.  We found that loss leveled out around 25,000 steps and stopped optimization at that point.  Upon convergence, the network was able to satisfy 94.504\% of the test triplets.
Figure~\ref{fig:ZapposEmbedding} shows samples of the Zappos50K data set embedded in two dimensions using the $\phi$ triplet network.


\end{document}